\documentclass{iau}
\usepackage{graphicx} 
\usepackage{url}

\title[IAUS291.~~RRAT discoveries in GBT Drift-scan Survey] %% short title %%
{Discoveries of Rotating Radio Transients in the 350\,MHz Green Bank Telescope Drift-scan Survey} %% full title %%

\author[Karako-Argaman \etal\ ]  %% short author list %%
{Chen Karako-Argaman$^1$
 \and the GBT Drift-scan Collaboration$^2$}

\affiliation{$^1$Department of Physics, McGill University \\ 3600 University Street, Montreal, QC, Canada, H3A 2T8 \\ email: {\tt karakoc@physics.mcgill.ca} \\[\affilskip]
$^2$ \url{http://www.as.wvu.edu/~pulsar/GBTdrift350/}}

\pubyear{2012}
\volume{291}  %% insert here IAU Symposium No.
\jname{\mbox{Neutron Stars and Pulsars: Challenges and Opportunities after 80 years}}
\editors{J. van Leeuwen, ed.} 
\begin{document}

\maketitle

%% -- Abstract ----------------------------------
\begin{abstract}
Rotating Radio Transients (RRATs) are a class of pulsars characterized by sporadic bursts of radio emission, which make them difficult to detect in typical periodicity-based pulsar searches. Using newly developed post-processing techniques for automatically identifying single bright astrophysical pulses, such as those emitted from RRATs, we have discovered approximately 30 new RRAT candidates in data from the Green Bank Telescope 350 MHz drift-scan survey. A total of 6 of these have already been confirmed and the remainder look extremely promising. Here we describe these techniques and present the most recent results on these new RRAT candidates.
\keywords{stars: neutron, pulsars: general, surveys}
\end{abstract}

% add below any authors, subjects and objects for indexing 
%   add more lines if necessary
%   but leave all lines commented out
%\index[author]{Karako-Argaman, C.|textbf}
%\index[subject]{pulsars:general}
%\index[subject]{surveys}

\firstsection % if your document starts with a section,
\section{Introduction}

\begin{figure}[t]
 \begin{center}
  \includegraphics[width=2.95in]{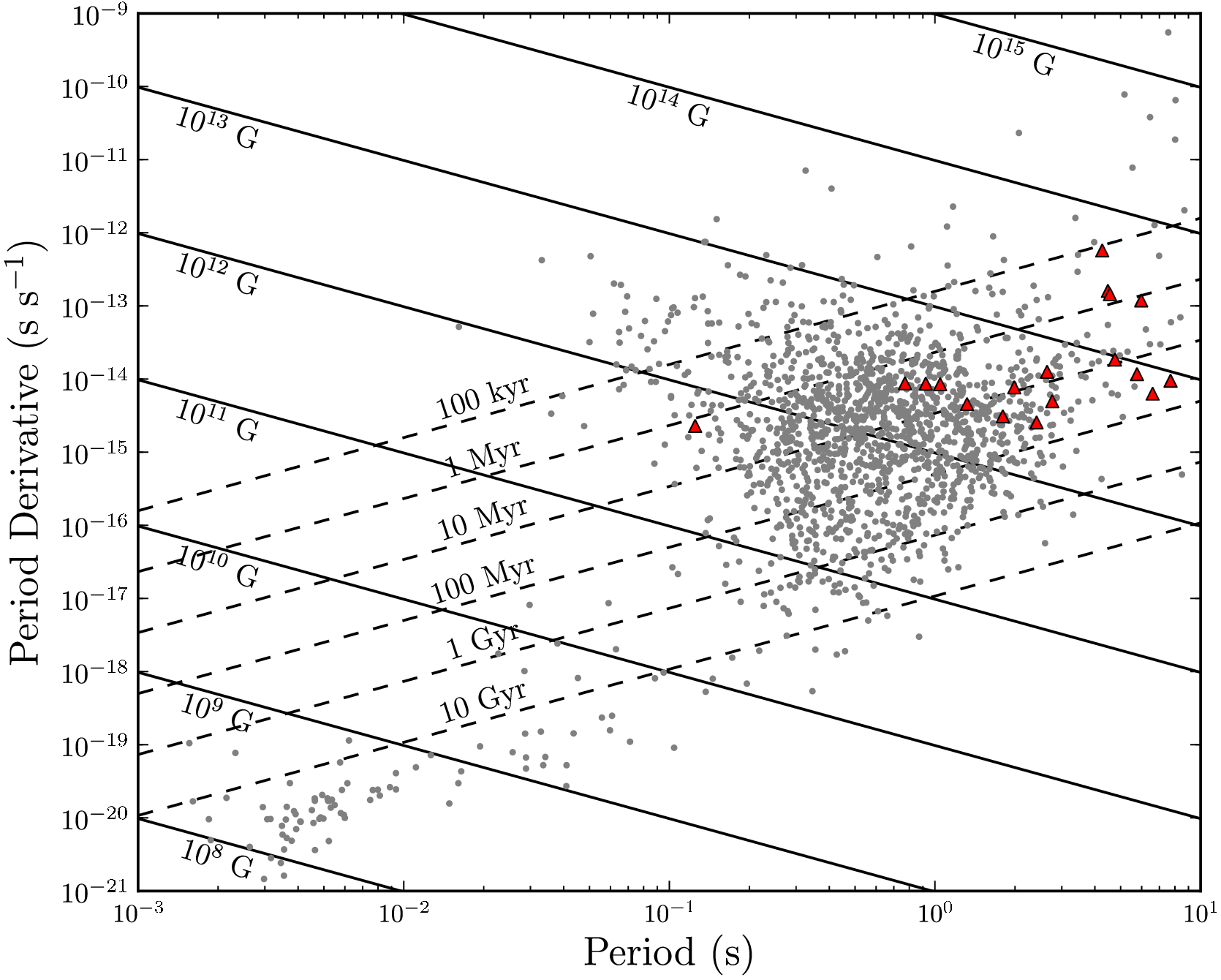}
  \caption{$P-\dot{P}$ diagram for all known neutron stars outside of globular clusters. The 18 RRATs with measured $\dot{P}$ are indicated by triangles, and dots represent other neutron stars. Lines of constant magnetic field are solid and lines of constant characteristic age are dashed. As can be seen, these 18 RRATs tend to have longer periods and larger magnetic fields than the bulk of long-period pulsars.} 
    \label{fig:PPdot}
 \end{center}
 \end{figure}

Rotating Radio Transients (RRATs) were first discovered by McLaughlin \etal\ \cite[(2006)]{McLaughlin_etal06} and are characterized by occasional single radio pulses, with no underlying periodicity easily detected, in contrast to typical rotation-powered pulsars.
The discovery of RRATs was surprising and important as it revealed a class of neutron stars that had previously been missed by pulsar surveys, thereby suggesting a potentially enormous increase in the inferred neutron-star birthrate.  Moreover the discovery raised the issue of why some neutron stars show such sporadic emission, whereas others are `on' continuously. RRATs may exhibit up to $\sim$1000 periods of separation between detected pulses, and as such are not detected in periodicity searches.
Furthermore, observational biases against detecting RRATs suggest that these sources represent a significant fraction of radio-active Galactic neutron stars, yet the current known population is very small.
Today $\sim$70 RRATs are known\footnote{See the ``RRATalog'' at \url{http://www.as.wvu.edu/~pulsar/rratalog/} for a list.} however the vast majority do not yet have their basic properties measured. Those RRATs whose properties have been measured tend to have longer periods and are mostly seen to lie between the regular pulsars and the magnetars, as shown in the $P-\dot{P}$ diagram in Figure~\ref{fig:PPdot}. This may suggest an evolutionary link between RRATs and other pulsar classes, but with such a small known RRAT population, it is difficult to draw conclusions regarding their place within the global pulsar population.

Since the initial discovery of the first 11 RRATs, there has been a major paradigm shift in radio pulsar searches. Single-pulse searches are now routine and a large number of new RRAT discoveries are being made (e.g. \cite[Keane \etal\ 2010]{Keane_etal10}; \cite[2011]{Keane_etal11}).
On the other hand, a lot of human involvement is required in examining the output of single-pulse searches in order to look for RRAT-like patterns.
This, along with the already present biases against finding RRATs, makes expanding and studying the RRAT population a difficult task. Hence, it is imperative to make automated RRAT searches a routine part of radio pulsar searching.

We have developed new post-processing techniques for automatically identifying single bright astrophysical pulses, such as those emitted from RRATs. In implementing this search algorithm on data from the Green Bank Telescope 350 MHz Drift-scan Survey, we have discovered 33 new RRAT candidates, of which 6 have already been confirmed. Here we describe these techniques and present our discoveries.

\section{The Green Bank Telescope 350 MHz Drift-scan Survey}
We have searched for RRATs in data from the Green Bank Telescope (GBT) 350 MHz Drift-scan Survey. This survey was conducted over the summer of 2007, while the azimuth track of the GBT was undergoing repairs, and thus the telescope was stationary at several fixed elevations, collecting data as the sky drifted over it. The data were analyzed in 140-s sections, corresponding roughly to the time it takes a point on the sky to pass through the full width half maximum of the telescope beam. This survey produced 134 TB of data, covering  over 10,000 square degrees of the sky, and has yielded 34 pulsars thus far. For more information on this survey, see \cite{Boyles_etal12} and \cite{Lynch_etal12}.

These data were then processed using the {\tt PRESTO} software suite (\cite{Ransom_01}), which included radio frequency interference (RFI) excision, de-dispersion, searches for periodic signals in the Fourier domain, and single-pulse searches. In single-pulse searches, signals in the de-dispersed time series that deviate from the mean significantly are identified, and information such as signal-to-noise, time, dispersion measure (DM), and pulse width is recorded. Diagnostic ``single-pulse plots" summarizing this information are then produced for each beam and saved for human inspection. In this survey, the data were divided into 30,000 140-s ``pointings", producing 120,000 such diagnostic plots (4 plots per pointing, each plot spanning a different DM range) that then required visual examination. This is a tedious, time-consuming task, and can thus be a bottleneck in discovering single-pulse sources.

\section{RRAT search algorithm}
\label{algorithm}
We have developed an automated search algorithm in order to identify RRAT candidates in the output of single-pulse searches described above, eliminating the need for manual inspection of each diagnostic plot produced.
Our algorithm is based on the following concepts:
\begin{enumerate}
  \item{\label{dmrange} A bright signal will be detected over a range of DMs, with the strongest detection at the optimal DM and weaker detections above and below this DM.} 
  \item{\label{rfidm0} Since signals are strongest at the optimal DM, we expect that signals of terrestrial origin (namely RFI) will peak at a DM of 0. We can thus classify any signals that peak at DM$\sim$0 as not astrophysical and reject them.}
\end{enumerate}
Concept (\ref{dmrange}) means that a given pulse, whether astrophysical or not, will be associated with many statistically significant ``single-pulse events" that will be found in the single-pulse search. These events will be spread over a small range of DMs, and will occur at approximately the same time. The first step in our algorithm is thus to {\bf group} events that belong to the same pulse by checking whether they satisfy this criterion, that is, lie within some small window of DM and time. Once the single-pulse events in a beam have been divided into groups, we {\bf examine each group's collective properties} in order to decide whether it behaves like an astrophysical pulse, and {\bf rate} it based on these results.

The first test we employ is group size. If a group has too few events, we classify it as noise. The following test examines signal-to-noise vs. DM behaviour. From concept (\ref{rfidm0}), we expect that a group of events that is due to an RFI signal will have a peak signal-to-noise at DM$\sim$0. Thus, any groups that satisfy this criterion are classified as RFI. Finally, we again make use of concept (\ref{dmrange}) by looking for groups that have peak signal-to-noise at some given (non-zero) DM, which then decreases above and below that given DM. These groups are classified as likely astrophysical pulses, and depending on the peak signal-to-noise attained, are classified as ``likely" to ``very likely" pulses.

Once the beams have all undergone this search algorithm, those that have been flagged as having ``very likely" pulses are visually examined. For those beams that indeed look like astrophysical sources, we then generate ``waterfall plots", which are frequency vs. time plots that show arrival times of the signal in different frequency bins. Since astrophysical signals experience a frequency-dependent dispersive delay while propagating through the interstellar medium, we use these waterfall plots as a final means of testing the astrophysical nature of a signal, before deciding whether it is a candidate worthy of follow-up.

\section{Results}
\begin{figure}[t]
 \begin{center}
  \includegraphics[width=2.5in]{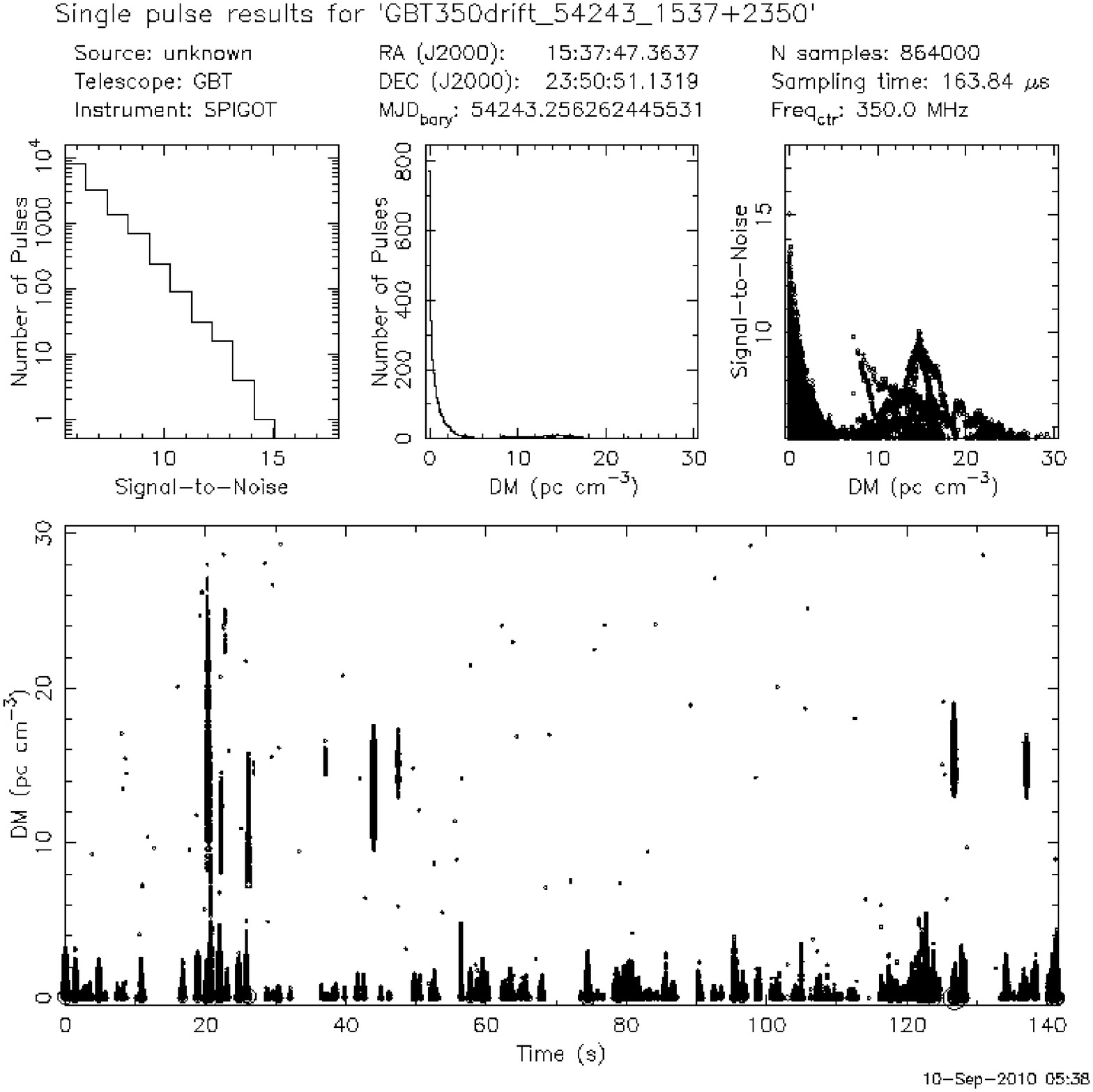}
  \includegraphics[width=2.5in]{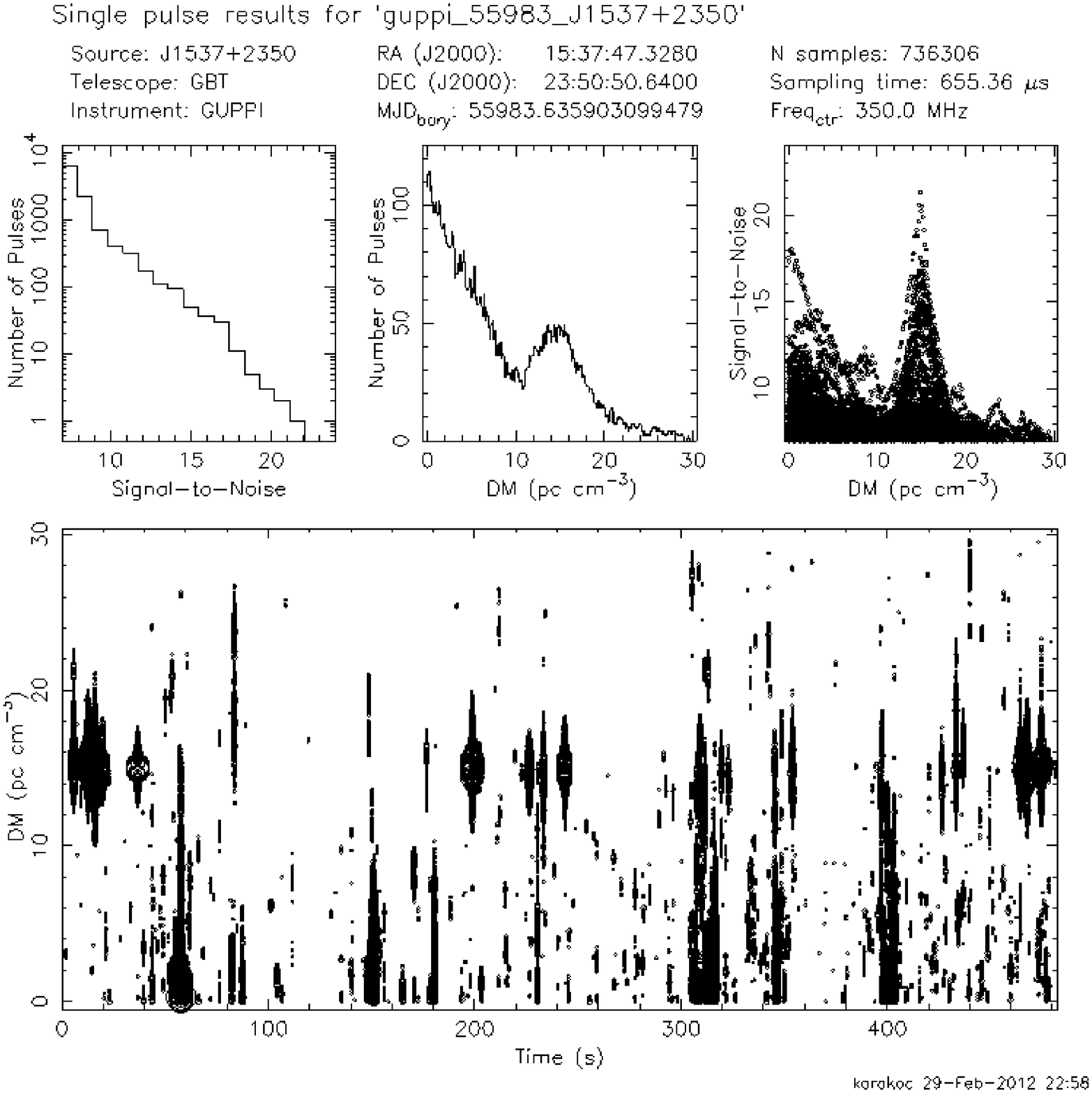}
  \caption{Discovery (left) and confirmation (right) single-pulse plots for newly discovered RRAT J1537+2350. The lower panel shows single-pulse events at the DM and time for which they were detected, with dot size proportional to signal-to-noise ratio. Notice, on the left, 5 pulses at DM$\sim$15~pc~cm$^{-3}$, starting at t$\sim$37~s, and various RFI spikes throughout. On the right, many strong pulses are seen, again at DM$\sim$15~pc~cm$^{-3}$. The top right panels of each diagram demonstrate the signal-to-noise vs. DM behaviour expected for an astrophysical pulse, as described in Section~\ref{algorithm}.}
    \label{fig:1537}
 \end{center}
 \end{figure}

We have processed all 30,000 pointings of the GBT Drift-scan Survey
with the search algorithm described above, resulting in a total of 33
RRAT candidates. Of the 7 candidates that have been followed up, 6
have been confirmed. The remaining sources look extremely promising,
based on their strong single-pulse detections, as well as dispersive
behaviour shown in waterfall plots. It should be noted that,
remarkably, the number of RRAT candidates found is comparable to that
of regular pulsars (34) found in the Drift-scan
Survey. Figure~\ref{fig:1537} shows an example discovery and confirmation plot of one of the new RRAT sources. The strength of the algorithm is illustrated in these plots: the pulses in the discovery observation may easily be confused with RFI when examined by eye, and this beam would have potentially been dismissed if it were only examined visually. The algorithm, however, successfully distinguished between the astrophysical pulses and the RFI in this beam, and correctly identified this source as an RRAT candidate.

\section{Future work}
We hope to begin timing the confirmed RRAT sources soon, allowing us to measure their basic parameters and place them on the $P-\dot{P}$ diagram. We will also follow up the remaining candidates, and begin timing observations of those that are confirmed.

We are also working on making improvements to the search algorithm that will allow us to identify weaker pulses in data, as well as make the algorithm more robust to RFI in order to reduce the number of false positives. Finally, since this code is not specific to the Drift-scan Survey it may be applied to other pulsar surveys. Indeed, in the coming weeks it will be implemented in the ongoing Green Bank North Celestial Cap Survey. Applying the search algorithm to this extensive sky survey promises to yield many more RRAT sources, since this survey will have improved sensitivity and will cover an area far larger than that of the GBT Drift-scan survey.

We hope that with this new automated search algorithm, the known RRAT population will quickly grow, allowing us to study more of these fascinating objects and find their place within the global pulsar population.


\begin{thebibliography}{}

\bibitem[McLaughlin \etal\ (2006)]{McLaughlin_etal06}
{McLaughlin, M.~A., Lyne, A.~G., Lorimer, D.~R., \etal\ } 2006, \textit{Nature}, 439, 817

\bibitem[Keane \etal\ (2010)]{Keane_etal10}
{Keane, E.~F., Ludovici, D.~A., Eatough, R.~P., \etal\ } 2010, \textit{MNRAS}, 401, 1057

\bibitem[Keane \etal\ (2011)]{Keane_etal11}
{Keane, E.~F., Kramer, M., Lyne, A.~G., \etal\ } 2011, \textit{MNRAS}, 415, 3065

\bibitem[Boyles \etal\ (2012)]{Boyles_etal12}
{Boyles, J., Lynch, R.~S., Ransom, S.~M., \etal\ } 2012, submitted to \textit{ApJ}, arXiv:1209.4293

\bibitem[Lynch \etal\ (2012)]{Lynch_etal12}
{Lynch, R.~S., Boyles, J., Ransom, S.~M., \etal\ } 2012, submitted to \textit{ApJ}, arXiv:1209.4296

\bibitem[Ransom 2001]{Ransom_01}
{Ransom, S. M.} 2001, PhD thesis, Harvard University

\end{thebibliography}
\end{document}